# Jitter analysis of a superconducting nanowire single photon detector

Lixing You[1], Xiaoyan Yang[1,2], Yuhao He[1,2], Wenxing Zhang[1,2], Dengkuan Liu[1,2], Weijun Zhang[1], Lu Zhang[1], Ling Zhang[1,2], Xiaoyu Liu[1], Sijing Chen[1,2], Zhen Wang[1], Xiaoming Xie[1]

[1] State Key Laboratory of Functional Materials for Informatics, Shanghai Institute of Microsystem and Information Technology, Chinese Academy of Sciences, Shanghai 200050, P. R. China
[2] Graduate University of the Chinese Academy of Sciences, Beijing 100049, P. R. China

Email: lxyou@mail.sim.ac.cn

Abstract:

Jitter is one of the key parameters for a superconducting nanowire single photon detector (SNSPD). Using an optimized time-correlated single photon counting system for jitter measurement, we extensively studied the dependence of system jitter on the bias current and working temperature. The signal-to-noise ratio of the single-photon-response pulse was proven to be an important factor in system jitter. The final system jitter was reduced to 18 ps by using a high-critical-current SNSPD, which showed an intrinsic SNSPD jitter of 15 ps. A laser ranging experiment using a 15-ps SNSPD achieved a record depth resolution of 3 mm at a wavelength of 1550 nm.

## 1. Introduction

Since the pioneering work of Gol'tsman in 2001, the superconducting nanowire single photon detector (SNSPD) has been proven to be the one of the best single photon detectors at near-infrared (NIR) wavelengths [1,2]. Especially in the past two years, system detection efficiency (SDE) has been effectively improved to over 70% [3,4,5], which shows affirmative prospects for future application. The SNSPD has a few extra parameters besides SDE, which are also important in various applications, for example, the dark count rate, repetition rate, and jitter.

Jitter of an SNSPD denotes the timing deviation from an ideal periodic single-photon-response voltage pulse, and it is crucial to applications such as light detection and ranging as well as high-speed quantum communication. For a typical Gaussian distribution, the jitter can be quantified using the full width at half maximum (FWHM) of the distribution. The SNSPD has been proven to have a lower jitter than traditional semiconducting SPDs at NIR wavelengths, which could give a better depth resolution for NIR laser ranging and imaging applications [6,7,8]. However, various values of system jitter from a few tens to around a couple of hundred picoseconds have been reported in the literature [3,6,7,8,8,9]. The spread of system jitter makes it difficult to determine the origin of the intrinsic jitter of a device quantitatively, and the mechanism of this intrinsic jitter is still unclear. The variation of the hot-spot size and/or location caused by photon absorption can contribute to the jitter [10].

A time-correlated single photon counting (TCSPC) system was set up to accurately measure the intrinsic jitter of an SNSPD. The jitter of each part in the system was evaluated, and the system was optimized to decrease the system jitter. By studying the jitter dependence on the bias current of the SNSPD at different temperatures, we noticed that the SNSPD jitter is related to the absolute bias current regardless of the critical current of the SNSPD. We further demonstrated that the signal-to-noise ratio (SNR), which is proportional to the absolute bias current, directly influences the SNSPD jitter. By introducing an SNSPD with a higher SNR, the jitter of the TCSPC system was reduced to a low value of 18 ps, which gave a depth resolution of better than 3 mm in a laser ranging experiment. The intrinsic jitter of the SNSPD was estimated to be 15 ps.

## 2. SNSPD and TCSPC system

SNSPD devices were fabricated from superconducting ultrathin NbN films on MgO substrates with a traditional meander structure using E-beam lithography and reactive ion etching. The lateral dimension of the optically sensitive area was 10-15 $\mu$m, and the nanowire linewidth was around 100 nm with a filling ratio of 0.5. Photons were irradiated onto the SNSPD through a single-mode fiber. Without using any optical cavity, the detector can have a typical system detection efficiency of 1–10% when it is installed in a Gifford-McMahon cryocooler and cooled down to 2.2 K.

For measuring the jitter of the SNSPD, a TCSPC system was set up (schematic in Figure 1(a)). A femtosecond pulsed laser (Calmar, FPL-01CAF) was adopted as the single photon source. The power of the photon pulse can be attenuated to the level of a single photon per pulse by an attenuator with a variable attenuation ratio. The synchronization signal of the laser was sent to the "Start" port of a commercial TCSPC module (Becker-Hickl, SPC-150) through an inverter. Because a typical single-photon-response voltage pulse has an amplitude of about 1 mV, it must be amplified in the read-out circuit. The amplified voltage signal was sent to the "Stop" port of SPC-150. SPC-150 builds a statistical distribution of the intervals between the "Start" and the "Stop" signals. For the SNSPD, the distribution typically shows a Gaussian profile, from which the system jitter can be obtained.

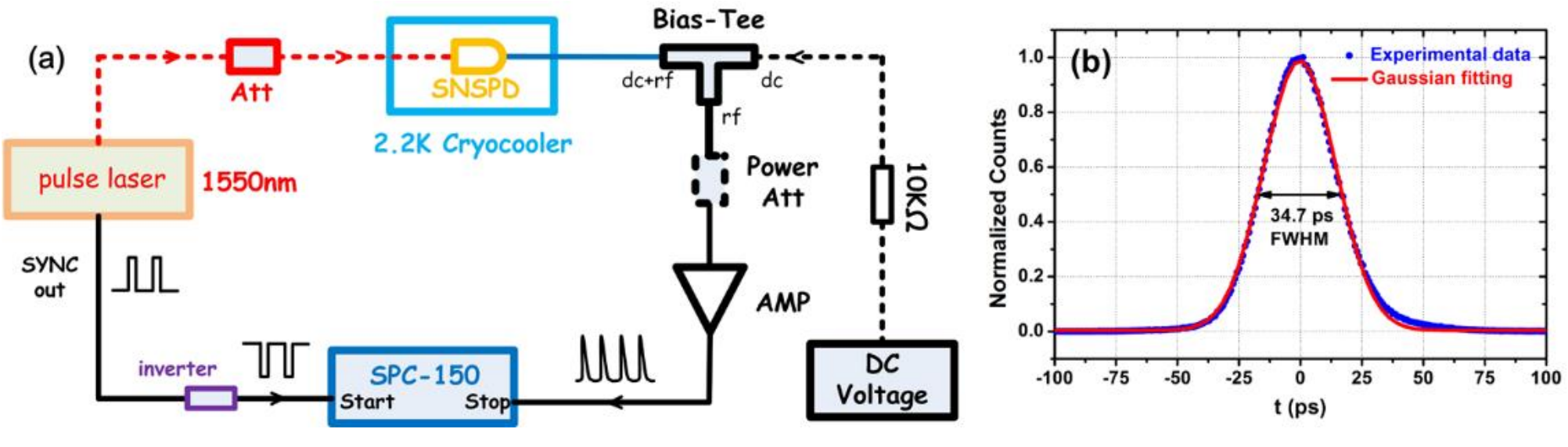

Figure 1. (a) Schematic view of the TCSPC system for measuring the jitter of the SNSPD. The power attenuator between the bias-tee and the amplifier is used in the experiment for tuning the SNR of the response signal; (b) a typical distribution of the system jitter $j_{system}$ = 34.7 ps, and the Gaussian fitting curve.

3. Jitter measurement and discussions

Each component in the TCSPC system can contribute to the system jitter $j_{system}$, which can be represented by [3]

$$j_{system} = \sqrt{j_{SNSPD}^2 + j_{laser}^2 + j_{SYNC}^2 + j_{SPC}^2}\,, \tag{1}$$

where $j_{SNSPD}$, $j_{laser}$, $j_{SYNC}$, and $j_{SPC}$ are the jitters from the SNSPD including the circuits, the laser, the synchronization signal of the laser, and SPC-150 respectively. Since the amplifier performance is related to the frequency spectrum characteristics of the single-photon response, it is somewhat difficult to separate the jitters from the SNSPD and the rest of circuit. All the components were carefully selected and optimized to decrease their jitters so we could obtain a lower $j_{system}$. The best results we obtained were 0.1 ps, 4.0 ps, and 7.6 ps for $j_{laser}$, $j_{SYNC}$, and $j_{SPC}$, respectively. Those small jitters contribute only a small portion to the system jitter, which ensures that $j_{SNSPD}$ can be deduced from Eq. (1) accurately. Figure 1(b) shows typical measurement data and the Gaussian fitting curve of the system jitter from which $j_{SNSPD}$ = 34.7 ps was obtained.

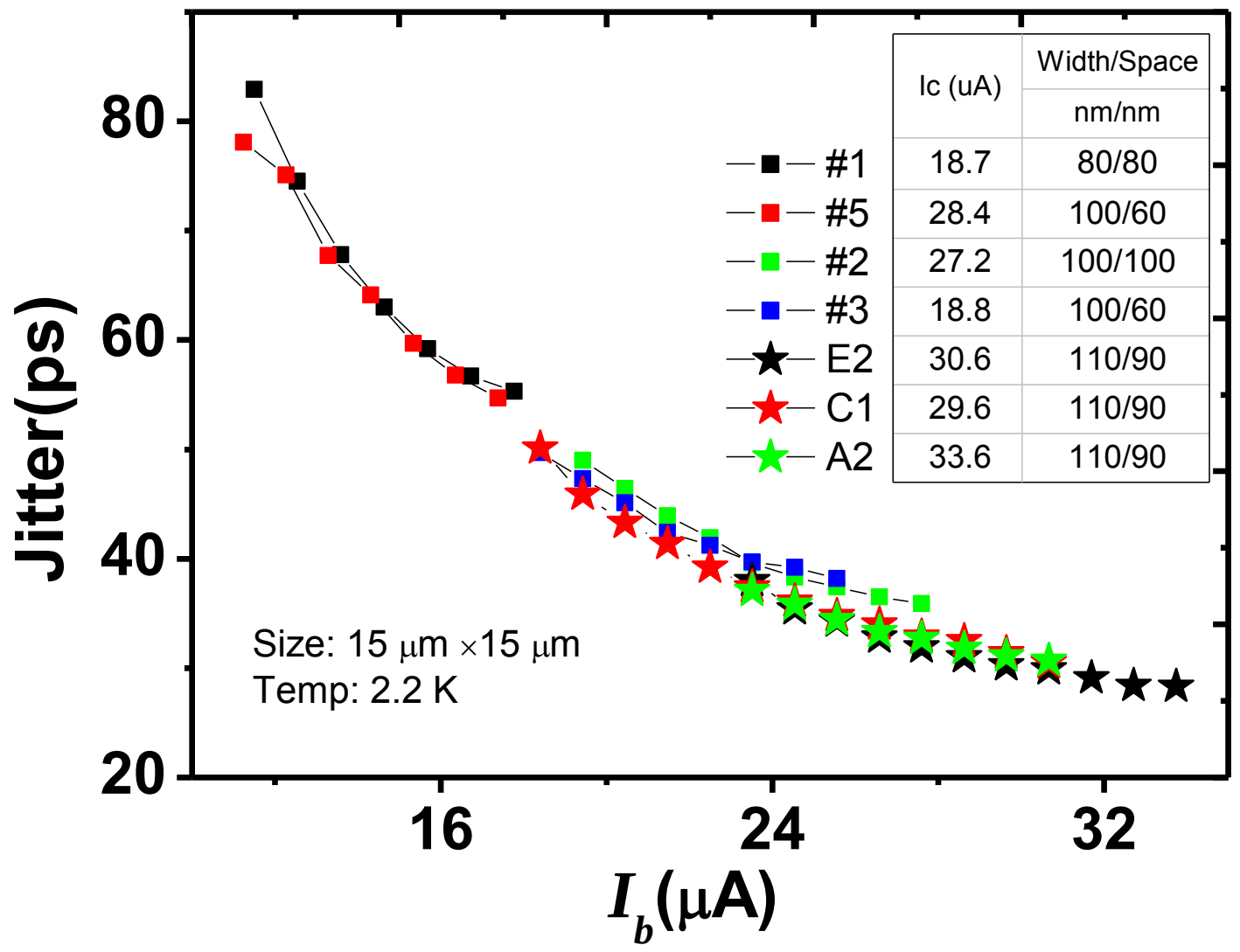

Figure 2. System jitter dependence on the bias current of different SNSPDs. All SNSPDs are fabricated from 5-nm-thick NbN film. The parameters of the critical current and the width/space of SNSPDs are shown in the inset.

To understand the origin of $j_{SNSPD}$, we studied jitter dependence on the bias current $I_b$. The jitter decreases with an increase in the bias current, which is consistent with results reported elsewhere [3,8,9]. Another more interesting result was observed for a few detectors fabricated in separate batches with different design parameters. Though the different SNSPDs have different critical currents $I_C$, values of $j_{system}$ ($j_{SNSPD}$) were nearly the same for the same bias current as shown in Figure 2. This result indicated that $j_{SNSPD}$ is related to the bias current of the SNSPD. Since the critical currents of those devices are different from each other, the relative bias currents ($I_b/I_C$) are different. To determine further whether $j_{SNSPD}$ is related to the absolute bias current or to the relative bias current, the jitter dependence on the bias current was measured at elevated temperatures, which affects the critical current of the device. The results are shown in Figure 3. We noticed that for the same device, though the critical current changed effectively, the jitter remained same for the same bias current, which further proves that $j_{SNSPD}$ is related to the absolute bias current of the SNSPD.

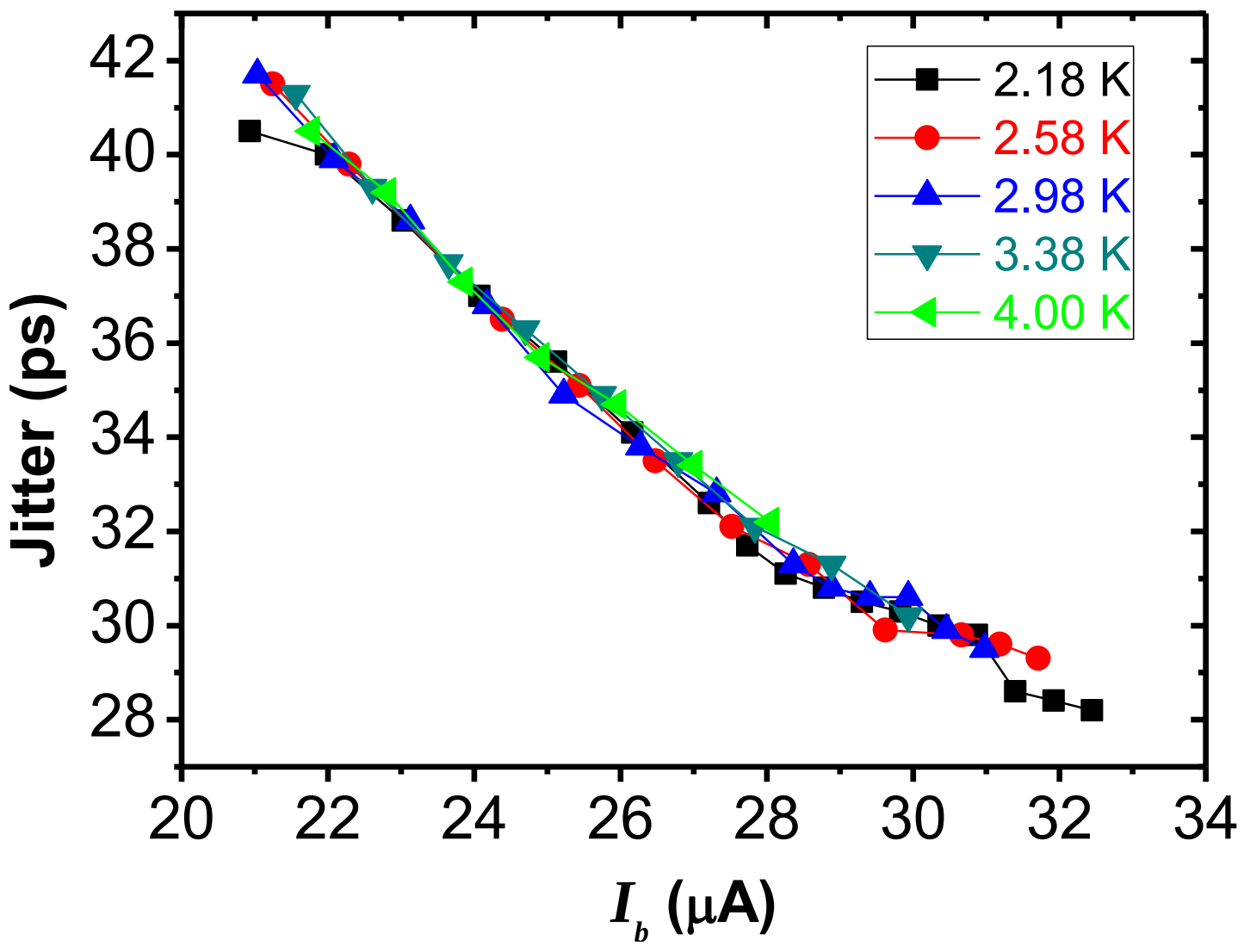

Figure 3. System jitter dependence on the bias current of Sample E2 at different temperatures. The temperature is controlled by heating the cold stage using a resistive heater.

Because the amplitude of the photon response signal is proportional to the absolute bias current of the SNSPD, the jitter dependence on the absolute bias current gives us a clue that the SNR may be the internal factor that influences the SNSPD jitter. With the same noise caused by the amplifier, a higher pulse amplitude results in a higher slope of the rising edge of the response pulse, i.e., a higher SNR, which produces less jitter due to the noise. To verify the SNR influence on the jitter, a tunable power attenuator was inserted into the circuit between the bias-tee and the amplifier (see Figure 1), which gave us the ability to change the SNR while keeping the same absolute bias current for the SNSPD. Figure 4 shows the jitter dependence on the SNR, or $\sigma_n/k$, where $\sigma_n$ and $k$ are the root-mean-square (RMS) noise and the largest slope of the rising edge of the response pulse respectively. The square dots in Figure 4 indicate that by decreasing the SNR, jitter increases even if the absolute bias current of the SNSPD remains constant. Further, we removed the power attenuator and instead changed the absolute bias current of the SNSPD, so that the SNR was adjusted to the same level as when we had used the power attenuator. The jitter results are shown as round dots in Figure 4. As long as the SNR is the same, the jitter remains constant.

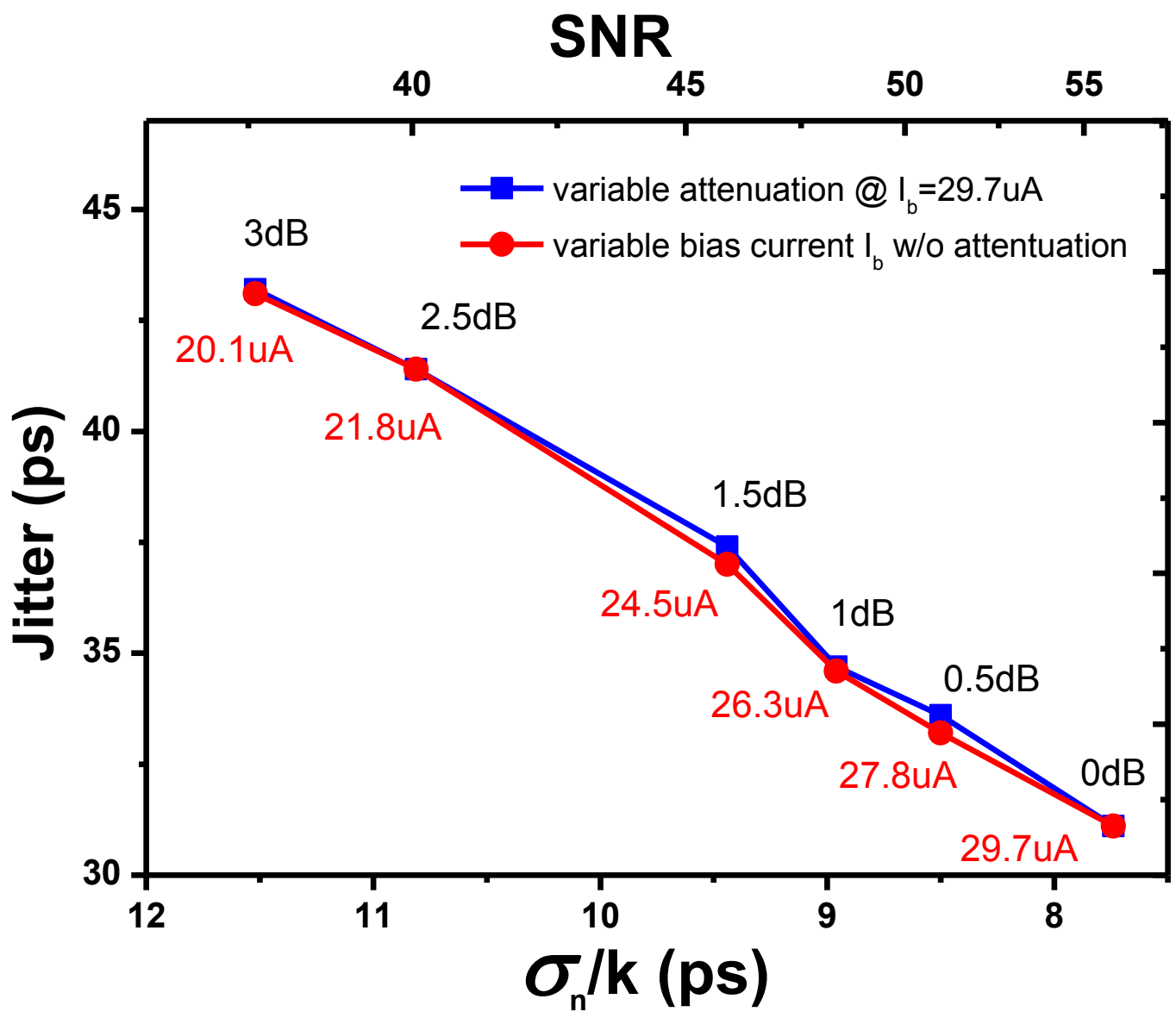

Figure 4. Dependence of system jitter on the SNR, or $\sigma_n/k$. The SNR or $\sigma_n$/k is controlled by changing either the attenuation ratio (square dots) or the bias current (round dots).

The above results prove that the SNR plays an important role in the SNSPD jitter. Supposing the jitter caused by the low SNR is independent, the system jitter can be rewritten as

$$j_{system} = \sqrt{j_{intr}^2 + j_{SNR}^2 + j_{laser}^2 + j_{SYNC}^2 + j_{SPC}^2}\ , \quad (2)$$

where $j_{intr}$ and $j_{SNR}$ represent the intrinsic jitter of the SNSPD and the jitter due to the low SNR, respectively. As a result, it is necessary to improve the SNR to obtain a smaller $j_{system}$. By choosing a low-noise low-temperature amplifier to decrease/increase the noise/SNR, the jitter can be partially reduced to some extent [3]. However, we might consider increasing the SNR by increasing the bias current (i.e., the amplitude of the response pulse) of the SNSPD. Driven by this idea, a specific batch of devices with a larger critical current were fabricated using 8-nm-thick NbN film (compared to the typical thickness of 5 nm) while keeping the nanowire linewidth with 100 nm. In this way, the critical current of the devices was increased from about 30 μA to more than 70 μA. Almost all the devices show a suppressed $j_{system}$ of about 18.0 ps.

Figure 5(a) shows a typical FWHM value of 18.0 ps. Moreover, we measured the jitter dependence of the bias current (60–72 μA) again (shown in Figure 5(b)). At lower bias current, it is difficult to measure the jitter due to the lower SDE. The jitter in Figure 5(b) saturated and did not change effectively as the bias current was increased, which indicates that $j_{SNR}$ has a negligible contribution to $j_{system}$. In fact, from the equation of the noise jitter $j_{SNR}$ = $2.355\sigma_n/k$ [11], we concluded that $j_{SNR}$ is less than 5 ps. As a result, the intrinsic jitter of the SNSPD $j_{intr}$ can be estimated roughly to be 15 ps from the above equation, which is the lowest result reported for an SNSPD with a size of 10 μm × 10 μm and is among the lowest results reported for all kinds of SNSPDs [5,10].

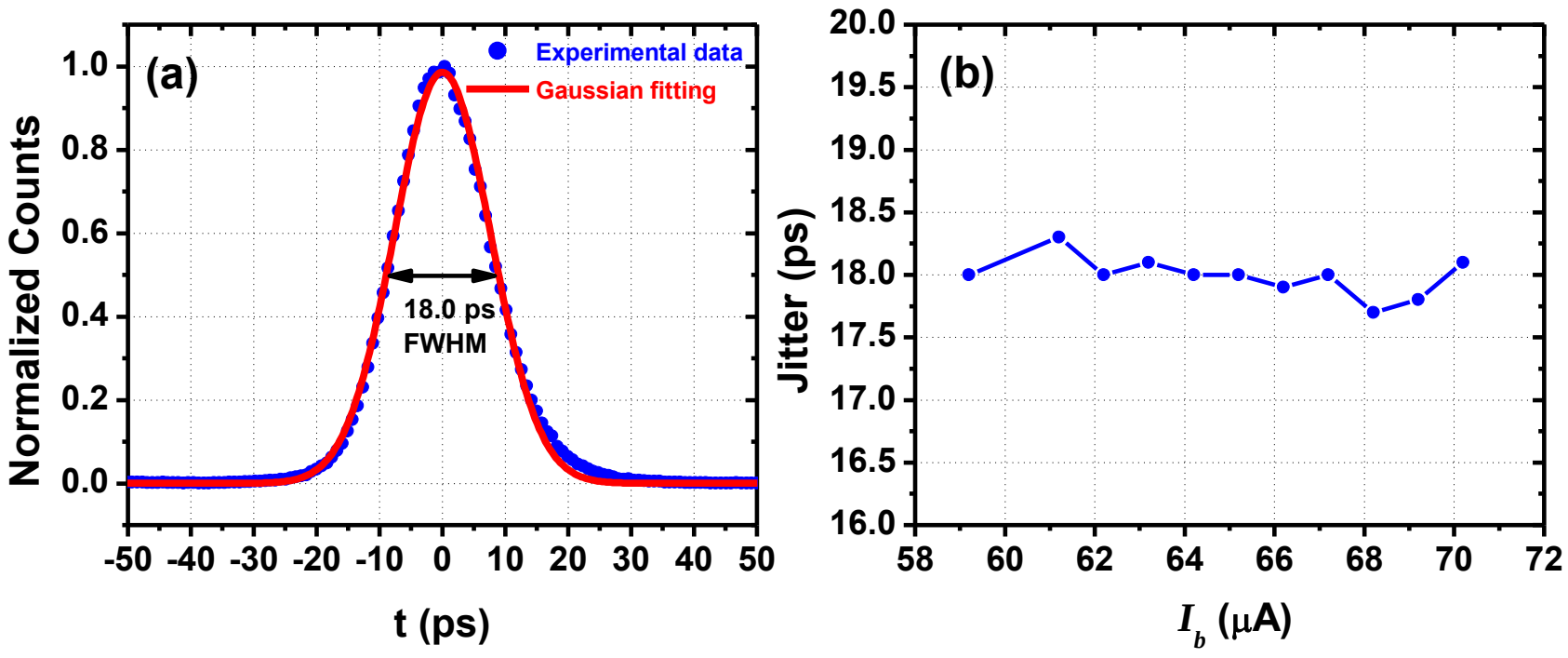

Figure 5. (a) System jitter of an SNSPD at a bias current of 64.2 μA; (b) system jitter dependence on high bias current.

4. Laser ranging experiment using the low-jitter SNSPD.

A low-jitter SNSPD can further improve depth resolution for laser ranging applications at 1550 nm. With a system jitter of 27.0 ps, we demonstrated time-of-flight (TOF) laser ranging at a working distance of 115 m. The depth resolution was 4.0 mm [8]. A similar experiment was conducted using a TCSPC system with a $j_{system}$ of 18.0 ps. The results are shown in Figure 6. A depth resolution of 3.0 mm was demonstrated, which is consistent with the theoretical expectation. To the best of our knowledge, this is the best result reported so far for the laser ranging at 1550 nm.

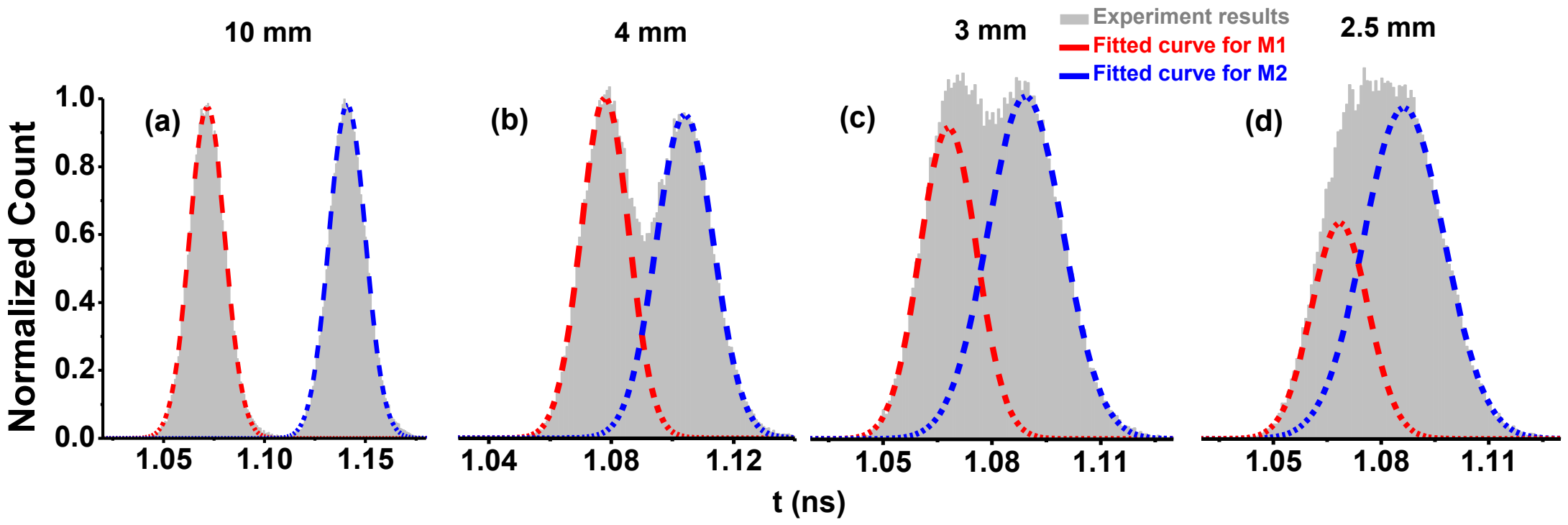

Figure 6. Laser ranging results using the low-jitter SNSPD. The experimental setup is described in Reference 8. (a)–(d): the distances between the two target mirrors (M1 and M2) are 10 mm, 4 mm, 3 mm, and 2.5 mm, respectively. The target distance is around 115 m.

5. Summary

The jitter of an SNSPD was extensively studied. We noticed that the jitter decreased with an increase of the absolute bias current rather than the relative bias current. The dependence of the jitter on the SNR was proved. By increasing the bias current of the SNSPD, the system jitter was effectively reduced to less than 20 ps, and the intrinsic jitter of the SNSPD was estimated to be 15 ps. A TOF laser ranging experiment was conducted using the low-jitter SNSPD, and a record depth resolution of 3 mm was demonstrated. The above results gave us a more accurate measurement of the intrinsic jitter of the SNSPD, which makes the quantitative experimental study of the origin of jitter possible in the future.

## 6. Acknowledgement

This work was supported by the National Natural Science Foundation of China (91121022), Strategic Priority Research Program (B) of the Chinese Academy of Sciences (XDB04010200 and XDB04020100), the National Basic Research Program of China (2011CBA00202), and the National High-Tech Research and Development Program of China (2011AA010802).

---